\documentclass{ws-ijmpd}

\newcommand{\Mpc}{\ensuremath{{\rm Mpc}}}
 
\newcommand{\mK}{\ensuremath{{\rm mK}}}

\newcommand{\MHz}{\ensuremath{\, {\rm MHz}}}

\begin{document}

\markboth{Xuelei Chen}
{21cm cosmology}

%
\catchline{}{}{}{}{}
%

\title{The Tianlai project: a 21cm cosmology experiment}

\author{Xuelei Chen}
\address{National Astronomical Observatories, Chinese Academy of Sciences\\
20A Datun Road, Chaoyang District, 
Beijing, 100012, China\\
xuelei@cosmology.bao.ac.cn}

\maketitle

\begin{history}
\received{Day Month Year}
\revised{Day Month Year}
\comby{Managing Editor}
\end{history}

\begin{abstract}
In my talk at the 2nd Galileo-Xu Meeting, I 
presented several different topics in 21cm cosmology 
for which I have done research. These includes
the 21cm signature of the first stars\cite{CM08,LZC07}, the 21cm signal 
from the IGM and minihalos\cite{YCSC09}, effect of dark matter annihilations
on 21cm signal\cite{YYBCZ10}, the 21cm forest by ionized/neutral 
region\cite{XCFTC09}, and the 21cm forest by minihalo and earliest 
galaxies\cite{XFKC10,XFC11}. In this conference 
proceeding I shall not repeat these discussions, but instead focus on  
the last part of my talk, i.e. the {\it Tianlai} project, an 
experiment effort on low redshift 21cm intensity mapping 
observation for dark energy measurements.
\end{abstract}


\section{Introduction}	

The ground state of the neutral hydrogen atom are hyperfine splitted by the 
interaction between the magnetic dipole of the eletron and proton. Photons
with 21cm wavelength (1420 MHz in frequency) 
are emitted or absorbed by the transition between these
hyperfine structure states. As the neutral hydrogen constitutes about 3/4 of 
the baryon matter in the Universe, these 21cm photons provide a very useful
tool for us to gather the information about the universe. Observations of the
21cm emission have long been used to map out the gas distribution in the 
Milky Way and the nearby galaxies. A very important evidence for the existence
of dark matter comes from the observation of rotation curves, and the 
21cm observation is one of the primary means of measuring the rotation 
curve.

During the last decade, there is an explosion of interest on 
21cm observation. This has been driven by developments in both science and
technology. With large amount of data coming from 
high precision cosmic microwave background (CMB) experiments,
large scale galaxy redshift surveys, type Ia supernovae, etc., an inflationary 
cold dark matter model with cosmological constant ($\Lambda$CDM) has emerged
as the concordance model of cosmology, which provides reasonably good and 
consistent description of most cosmological observations at a quantitative 
level\cite{sci03}. 
It then becomes interesting and plausible to investigate how the 
first stars and galaxies form in such a universe. With the discovery of the 
Gunn-Peterson trough in the redshift ($z>6$) quasars\cite{Fan02}, 
the epoch of reionization (EoR) 
begin to attract strong interest from observers and theorists 
alike. As most of the gas in the Universe is neutral before reionization, 
it is natural to consider the redshifted 21cm line as an important observational
probe of the EoR. At the same time, from a technological perspective, 
the developments in digital electronics leads to the concept of ``software
telescope'' which 
allows substantially larger and more precise interferometers to 
be built, and the 
digital signal processing means that one can remove foregrounds and extract
the 21cm signal even when the foregrounds are orders of magnitude greater than
the signal. Currently, several dedicated 21cm experiments, including 
the LOFAR in Europe(http://www.lofar.org), MWA in 
Australia (http://www.haystack.mit.edu/ast/arrays/mwa) and 
the 21CMA in China (http://21cma.bao.ac.cn) 
have started to take advantage of these developments and look
for signs of reionization.

The 21cm experiments draw much lesson from the CMB experiments. Unlike
previous radio astronomy observations where one trie to obtain high 
signal-to-noise maps of individual radio sources, in the CMB experiments 
low signal-to--noise ratios per pixel are standard fare, and instead of 
maps of single sources, statistical variables such as the angular power
spectrum plays the central role. This new approach greatly reduced the 
technical and financial demand on the experiment. 
Thus, all of the EoR 21cm experiments are designed to measure 
{\it statistics} of 21cm signal. 
It is then realized that the same principle, now dubbed 
as {\it intensity mapping}, can also be applied to experiments
at lower redshifts\cite{PBP06,P09}. A specific application would be 
observation of the {\it baryon acoustic oscillation} (BAO) features on the 
matter power spectrum \cite{CPPM08}. 

\section{Baryon Acoustic Oscillation}

The BAO are just sound waves in the baryon-photon fluid before the 
decoupling of photons from baryons at the epoch of 
Recombination. Due to the characteristic scale of cosmic expansion, it
left wiggles on the matter power spectrum. On large scales, the galaxies 
number density disitrbution trace the dark matter density distribution, so 
the same feature also appears on the galaxy power spectrum. 
Such features, with the first peak 
at the $\sim 100 h^{-1}\Mpc$ scale, have been detected in galaxy 
redshift surveys such as the SDSS\cite{E05} and 2dF\cite{C05}. 
Given a cosmological model, the absolute scale of the BAO features 
can be calculated, thus it may serve as a standard ruler in cosmology,
and using it one can measure the angular diameter 
distance $D_A(z)$ and the Hubble expansion rate $H(z)$, as a distance 
in the direction perpendicular/parallel to the line of sight is given by
\begin{eqnarray}
r_{\perp} &=&(1+z)D_A(z)\Delta \theta,  \nonumber \\
r_{\parallel} &=&\frac{ c \Delta z}{ H(z) }.
\label{eqn:staRuLer}
\end{eqnarray}
Observation of the scales at
different redshifts can then be used to determine the cosmological model
parameters. In a model with dark energy equation of state
$w(z)$, these are given by 
\begin{equation}
\label{eqa:HubZ}
\frac{H(z)}{H_0}=\left[\Omega_m(1+z)^3+  \Omega_k(1+z)^2 +
\Omega_X e^{3\int_0^z\frac{1+w(z)}{1+z}dz} \right]^{1/2}
\end{equation}
and 
\begin{equation}
\label{eqa:DaZ}
D_A(z)=\frac{c}{1+z}\int_0^z\frac{dz}{H(z)} .
\end{equation}
From such observations, the properties of dark energy and can
be measured. Indeed, in the Dark Energy Task Force (DETF) 
report\cite{DETF06}, 
the BAO method was recognized as one of the primary probes of dark energy.
The underlying physics of the BAO is relatively clear, so it allows 
better modeling and control of the systematic errors. 
At the same time, high statistical precision can be achieved with
surveys of large effective volumes. A number of experiments, such as the 
SDSS-III BOSS, the WiggleZ, the BigBOSS, and the LAMOST \cite{W09}, 
have been or are being planned to exploit this. 
Even more ambitous projects, such as the Sumira (WFMOS) on the Subaru telescope,
and the JDEM-ADEPT space telescope have also been considered. 

The same measurement can be made with the 21cm intensity mapping experiment. 
The 21cm emission from a large scale ``cell'' would be proportional to the 
matter density in that cell, so the fluctuation of the 21cm could also reveal 
the BAO feature. Recently, this has been demonstrated in principle by 
correlating the GBT observed 
21cm emission in a small patch of sky at $z=0.8$ with the 
optically detected galaxies in the Deep2 survey same region \cite{CPBP10}.

\section{Design of the experiment}
The BAO features are only ``wiggles'', its detection
requires high precision measurement of the matter power spectrum.
To beat down the cosmic variance in such measurement, very large volume is 
required. Essentially, one needs to observe a sizable fraction of the 
cosmic lightcone for which we can observe, hence large sky area is required.
It is also necessary to make the observation at different redshifts. To probe
the dark energy, the relatively low redshift range $0<z<2.5$ are perhaps most
useful, because the dark energy becomes dominate at relatively low redshifts. 

The primary difficulty in 21cm observation is foreground. The largest foreground
in the relevant frequency range is galatic synchrotron emission, which scales 
as $\nu^{-\alpha}$, where $\alpha \sim 2.5$. Other foregrounds include 
bremsstrahlung, point radio sources, etc. Fortunately, most foregrounds, though
much stronger than the 21cm signal, are very smooth in frequency, whereas the
redshifted 21cm signal varies in the frequency. Thus, if one looks in one 
direction, one can remove the smoothly varing part of the frequency spectrum, 
and recover the rapidly varying 21cm signal. 

However, thermal noise could also
produce apparant variations in frequency. It is therefore necessary to achieve
very low noise temperature. The noise temperature per pixel is given by 
\begin{equation}
\delta T = \frac{T_{sys}}{\sqrt{2\Delta\nu t}}
\end{equation}
where $\Delta\nu$ is the bandwidth, $t$ the integration time on that pixel,
and $T_{sys}$ is the system temperature, which is generally given by 
incoherent sum of the sky temperature and the noise temperature in the 
receiving system, $T_{sks}=T_{sky}+T_{rec}$. At the frequency of interest 
(0.4-1.4 GHz), the cool part of the sky (i.e. outside the galactic plane and 
strong radio point sources) the sky temperature is at the level of a few to 
a few tens K. If a well designed receiving system is used, we can expect
to achieve a total system temperature of $50-100$K. By contrast, the 21cm signal
expected is a few mK. Thus, to achieve $SNR \sim 1$ (i.e. $\delta T \sim \mK$)
for a single pixel, $\delta\nu t \sim 10^{10}$. If we require 
$\Delta\nu \sim 0.1 \MHz$, the integration time is about $10^5 s$ per pixel,
which is about one day integration time per pixel. Only small patches of sky
could be observed with existing radio telescopes at this level.

If one wants to build a dedicated telescope for the 21cm intensity mapping 
observation, how should one proceed? Interferometer array would be a better
option than single dish, since the interferometer is more stable against 
variations in system gain, which is very important in the survey of large
scale structure. 

As one can see from the above equation, 
the brightness sensitivity can not be
improved with an increased in telescope aperture. Instead, one needs multiple
receivers, either on a single telescope, or on multiple telescopes, to increase
the total integration time. Note that one may still detect the 
BAO signal from a large number of pixels even if the individual pixel 
has SNR less than 1. To increase the 
number of pixels, one has to increase the observed area of sky. However, if
the observed area of the sky is increased at the expense of decreasing the 
integration time on individual pixels, the overall sensitivity can not 
be greatly improved. The instaneous field of view should also be large.

To resolve the BAO peaks, compact inteferometer arrays with longest base line 
of about a hundred meter seems to be a good compromise. At $z\sim 1$, this 
would give a best resolution of about 14 arcmin, corresponding to a comoving
scale of about $10 h^{-1}\Mpc$. Most baselines would be about half of this 
size, corresponding to scales twice large. This would provide sufficient
angular resolution to observe the higher BAO peaks in the matter power spectrum.

In Ref.~\refcite{PBP06}, radio telescope with cylinder design was proposed as
a possible way to build such an interferometer with low cost and fast survey
speed. The low frequency cylinder reflectors can be built with low cost, and 
it provides a way to realize large field of view observation.

\begin{figure}[t]
\centerline{\psfig{file=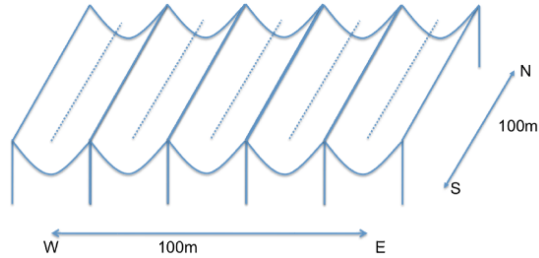,width=7.2cm}}
\vspace*{8pt}
\caption{\label{cylarray} The basic cylinder array design layout.}
\end{figure}

In Fig.~\ref{cylarray}, a tentative design of the array is shown. Five 
cylinder reflectors are arranged to lay side by side. The axes of the cylinders
are north-south oriented, with 15m-20m width and 100m length. 
Receiver feeds are packed along the focus line of the
parabolic cylinders, with a spacing of half wavelength to avoid multiple 
peaks in the synthesis beam profile. If we are to observe redshift 1, 
the half wavelength is about 21cm, so there would be about 500 dual 
polarization feeds along each cylinder\footnote{The dual polarization is 
essential, because the magnetic field in space cause frequency-dependent 
Faraday rotations of the polarized foreground radiation, this may induce 
a modulation in frequency if one observes only one polarization, mimicing the
21cm signal we search for. It is also necessary to calibrate the polarization
response with high precision to remove this.} 
The whole array would produce about 5000 signal inputs. For the convenience of
Fast Fourier Transform (FFT), we will choose to have $2^9=512$ feeds per 
cylinder. 

\begin{figure}
\centerline{\psfig{file=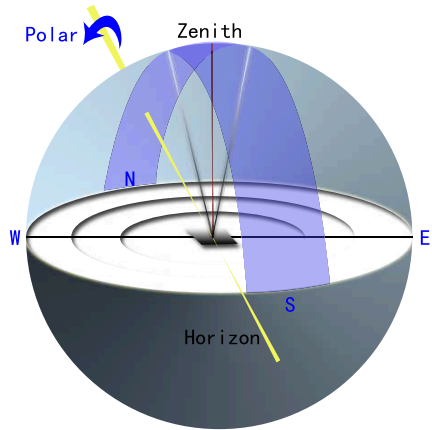,width=6.7cm}}
\vspace*{8pt}
\caption{\label{instview} The instaneous field of 
view of the telescope. As the Earth
rotates, the field of view scans across much of the sky.}
\end{figure}

The instaneous field of view is a narrow strip along the north-south direction
passing the zenith. In principle the strip may extend to horizon, if large
FOV feed is used in the cylinder. In practice, the strip would not extend that
far.

The cylinders shown in Fig.~\ref{cylarray} is of the on-axis 
design, it has the advantage of providing a symmetric primary 
beam. Alternatively, 
off-axis design can be used (see. e.g. Fig.1 in 
Ref.\refcite{S09}). The off-axis design has an unsymmetric primary beam in the
east-west direction, but allows easier access to the feeds.

\begin{figure}
\centerline{\psfig{file=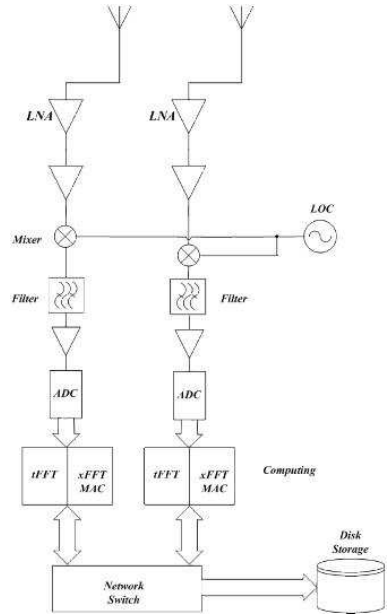,width=5cm}}
\vspace*{8pt}
\caption{\label{sigflow} Receiver Chain.}
\end{figure}

A schematic of the receiver chain is shown in 
Fig.~\ref{sigflow}. The incoming radio wave is reflected by the cylinder 
reflector, picked up by the feed, then amplified by the low noise 
amplifier (LNA). The local oscillator (LO) converts the radio frequency (RF)
band to intermediate frequency band (IF). The analog to digital converter (ADC)
converts the band-filtered analog signal to digital signal. In standard
FX correlator, one first makes an FFT of the time series data (tFFT), which 
reduces the computation from $O(N^2)$ to  $O(N\log N)$, 
then compute the cross correlations 
between the signals from different receivers
but of the same frequency. In the present case, the number of receivers 
are very large. To further reduce the amount of computation, one can also
make a further FFT on the spatial domain (xFFT). The data is 
exchanged through a network switch, redistributed so that at each computing 
unit, all the data needed for that FFT is available. 
For cylinder telescopes, the number of receivers is large along the 
cylinder, but the number of cylinders
is small, so spatial FFT along one dimension is sufficient. The cross
correlation between data on the 
same time and spatial frequency, across the different cylinders are computed
(multiply and accumulate, MAC). The results after integration of a time interval
of the order of seconds
(visibilities) are stored in harddrive for further analysis.

\section{Experiment Effort in China}

The National Astronomical Observatory of China (NAOC) has started 
research on the 21cm experiment for dark energy detection. We have
named it the {\it Tianlai} (cosmic sound) project \footnote{The word tianlai,
literally meaning the sky sound, comes from the work of ancient 
taoist philosopher Chuang-Tzu.}. We are collaborating with Jeff Peterson (CMU), 
Ue-Li Pen (CITA), Reza Ansari (U. Paris-Sud) and Kris Sigurdson (UBC)
in this research.

If we look at the history, we see that in building each telescope based 
on new technology, one would encounter unexpected problems. Only after these
problems are overcome could the new instrument work. As a first step to 
the full scale experiment, we can build a small scale prototype 
experiments in which the basic principle and different
designs of the experiment can be tested, and possible problems identified. 
Two small cylinders have been built at the Carnegie-Mellon University by 
Jeff Peterson and his collaborators. However, the site is not ideal as it is
located in central Pittsburgh with a lot of RFI.
The NAOC has decided to build three prototype cylinders at a quiet site to 
further the testing. 

The primary criterion for selecting the site is its electromagnetic  
environment. Site with low radio frequency interference (RFI) is desired.
The main source of RFI at this frequency range is mobile phone signal and 
TV broadcasting. The RFI can not be completely avoided, with high sensitivity
they would always appear in the EM spectrum. However, the RFI are typically
of narrow bandwidth, and as long as the EM field strength is not too strong
as to saturate or distort the output of the amplifier, it is possible to 
remove them with post processing. Of course, it is best to also have 
good logistic support, including road, electricity, and communication 
networking. The presence of existing astronomy or other science research
station could save our effort on logistic work.

A possible site of the experiment is in MingTuAn, ZhengXiangBaiQi in 
inner Mongolia, which is about 400km from Beijing. This is the site of the 
NAOC solar heliograph. We are also investigate other possible sites. 
We will make a selection after we have surveyed the RFI of these sites.

In the prototype experiment, 
we shall combine several inputs into one after the LNA. This 
would restrict the field of view, but would greatly reduce the required
digital electronics and the associated cost, while still allowing a good
test of the basic principles.

\section{Conclusion}

The 21cm experiment is very promising in cosmology, but it is also 
exceedingly hard. To detect the weak signal out of the huge foreground, 
it requires unprecedanted precision in calibration, not to mention 
the challenges in building a system with thousands of low noise receivers,
and process the data at real time. 
At present, no 21cm experiment has not yet made a positive detection. It
is still an uncharted, inviting virgin land. However, we believe that with
current technology, the experiment is possible, it just needs to be done.

\section*{Acknowledgments}

This work is supported 
by the NAOC, by the center of high energy physics of Peking University,  
by the John Templeton Foundation ``Beyond Horizon'' program in China, 
and by the NSFC grant No.11073024.


\end{document}